\documentclass[prd,onecolumn,preprintnumbers,amsmath,amssymb,superscriptaddress,showkeys]{revtex4-1}
\usepackage{bm,curves}
\usepackage{epsfig}
\usepackage{natbib}
\usepackage{dcolumn}
\usepackage{lineno}
\usepackage{amsmath}
\usepackage{slashed}
\usepackage{xcolor}
\usepackage{subfigure}
\usepackage[utf8]{inputenc}
\modulolinenumbers[5]

\def\lan{\langle }
\def\ran{\rangle }
\def\ksl{\slashed{k} }
\def\qsl{\slashed{q} }

\begin{document}

\preprint{NT@UW-17-15}

\title{Quantifying finite-momentum effects in meson quasi-PDFs}

\author{T. J. Hobbs}
\email{tjhobbs@uw.edu}
\affiliation{
        Department of Physics,
         University of Washington, Seattle, Washington 98195, USA
}
\begin{abstract}
The recently proposed large momentum effective theory (LaMET) of Ji \cite{Ji:2013dva} has led to a burst of activity among lattice
practitioners to perform and control the first pioneering calculations of the quasi-PDFs of the nucleon. These calculations
represent approximations to the standard PDFs defined as correlation functions of fields with lightlike
separation, being instead correlations along a longitudinal direction of the operator $\gamma^z$; as such,
they differ from standard PDFs by power-suppressed $1 \big/ p^2_z$ corrections, becoming exact in the
limit $p_z \to \infty$. Investigating the systematics of this behavior thus becomes crucial to understanding
the validity of LaMET calculations. While this has been done using models for the nucleon,
an analogous dedicated study has not been carried out for the $\pi$ and $\rho$ quark distribution functions.
Using a constituent quark model, a systematic calculation is performed to estimate the size and $x$ dependence of the finite-$p_z$
effects in these quasi-PDFs, finding them to be potentially tamer for lighter mesons than for the collinear quasi-PDFs of
the nucleon.
\end{abstract}


\date{\today}
\maketitle
%
\section{Introduction}
\label{sec:intro}
\paragraph{Motivation.}
The pion occupies a special position in the spectrum of light hadrons given its dual identities as the Goldstone boson associated
with the spontaneous breaking of chiral symmetry in QCD as well as the lightest meson with nontrivial quark content.
The fact that the pion is responsible for important contributions to the structure of the nucleon has been known for some time,
and the recognition of this fact has led to the introduction of pionic modes into many models \cite{Thomas:1981vc} of proton structure.
At the same time, ambiguities remain concerning the structure of the pion itself (as well as of other light mesons) and
its role in the nucleon's quark substructure and flavor asymmetries \cite{Peng:1998pa}. This fact has
inspired a number of experimental measurements of the pion structure function or parton distribution functions (PDFs) over the years,
including an approved JLab measurement of the $\pi$ structure function via spectator tagging in the Sullivan process \cite{Annand:2015}.
Theoretical studies of the $\pi$ structure function have an extensive history, with many model calculations of the $\pi$ structure function
making up an extensive literature \cite{Szczepaniak:1993uq,Frederico:1994dx,Bentz:1999gx,Hecht:2000xa,Ji:2004hz} on the topic.
Parallel to these efforts, the recently proposed large-momentum effective theory of Ji \cite{Ji:2013dva} has raised the prospect of
computing {\it quasi-PDFs}, which represent correlation functions of a spacelike operator $\gamma^z$ rather than the usual lightlike $\gamma^+$
that occurs in the formal definition of the {\it exact} PDF --- a fact which makes the latter undefinable on a Euclidean lattice as would be
required for lattice gauge calculations. These objects have thus enjoyed a rapidly burgeoning phenomenology
\cite{Chen:2016fxx,Radyushkin:2016hsy,Radyushkin:2017ffo,Carlson:2017gpk} in the past few years.
Crucially for the quasi-PDFs, as the longitudinal momentum $p_z$ of the parent hadron is boosted away from the rest frame, the
longitudinal separation over which quark fields are correlated is mapped onto the light front asymptotically. Effectively, sub-leading $p_z$
dependent corrections to the quasi-PDF are suppressed for $p_z \to \infty$ such that, in the limit of large hadronic momentum, LaMET provides a means
of computing the PDF with a precision limited only by the $p_z$-scale attainable in lattice calculations. In practice, however, efforts on the
lattice up to the present time \cite{Lin:2014zya,Chen:2016utp} have been confined to relatively small $p_z \sim 1$ GeV; this fact necessitates
careful study of the size and behavior of such corrections to the quasi-PDF due to the finite size of $p_z$ in likely lattice calculations.
While studies of this sort have been carried out for the nucleon using quark-diquark models \cite{Gamberg:2014zwa}, it is the goal of
the present work to use a constituent quark model to analyze the potential finite-$p_z$ effects in the valence-region quasi-PDF of
light mesons, specifically, the $\pi$ and $\rho$ meson.
Some analyses \cite{Radyushkin:2017gjd,Broniowski:2017wbr,Nam:2017gzm} of $\pi$ structure bearing on LaMET methods have recently emerged, but these have
been in the context of different model frameworks and formalisms, with the preponderance of the focus up to the present time on the parton distribution
amplitude (PDA), which is now being computed in the first lattice results \cite{Zhang:2017bzy}. Here, however, I undertake a concentrated analysis of the
$\pi$ and $\rho$ valence quasi-PDFs for the purpose of assessing the ability of the LaMET methodology to extract the high $x$ behavior of the meson
valence PDFs $q_\pi(x)$ and $q_\rho(x)$.
The remainder of this paper is as follows: Sect.~\ref{sec:formal} below describes in detail the theoretical framework --- the
method for computing the exact LF distributions of the $\pi$ and $\rho$ mesons as well as the corresponding calculations
using the LaMET method. In Sect.~\ref{sec:numerics} I describe my method for determining the model parameters that enter the
exact LF calculation as well as the size and $x$ dependence of the finite-$p_z$ corrections to the meson quasi-PDFs at several
assumed values of $p_z$. A brief conclusion follows in Sect.~\ref{sec:conc}.
%
%
%
%
\section{Formalism}
\label{sec:formal}
The universal parton distribution function (PDF) for a quark
carrying momentum $k$ in a hadron of momentum $p$ is given by Fourier transforms of hadronic matrix elements of 
parton-level correlation functions at a renormalization scale $\mu^2$, {\it viz.}
\begin{align}
\label{eq:LFform}
	q(x, \mu^2) &= \int {d\xi^- \over 4\pi} e^{-i \xi^- k^+} \lan p\, \big| \overline{\psi}(\xi^-) \gamma^+ \mathcal{U}(\xi^-,0)
	\psi(0)\, \big| p \ran \\
	\widetilde{q}(x, \mu^2, p_z) &= \int {d\xi_z \over 4\pi} e^{-i \xi_z k_z} \lan p\, \big| \overline{\psi}(\xi_z) \gamma^z \mathcal{U}(\xi_z,0)
	\psi(0)\, \big| p \ran\ +\ \mathcal{O} \left( {\Lambda^2 \over p^2_z}\, , {M^2 \over p^2_z} \right)\ ,
\label{eq:LMETform}
\end{align}
in which the quantities $\mathcal{U}(\xi,0)$ above represent Wilson line insertions in the longitudinal direction either along the light-front
or collinear to the $p_z$ boost, respectively. Following the work of Ji in Ref.~\cite{Ji:2013dva}, the second quantity in Eq.~(\ref{eq:LMETform})
above, $\widetilde{q}(x, \mu^2, p_z)$, is the {\it quasi-PDF}, which receives sub-leading corrections in $1/p^2_z$ that vanish in the limit
$p_z \to \infty$; this renders the LaMET framework an asymptotically accurate approximation of the exact LF PDF $q(x, \mu^2)$ in the limit of
infinite hadronic momentum. I verify this point numerically in the context of a constituent quark model for both the $\pi$ and $\rho$ systems,
with the goal of assessing the range and degree of validity of LaMET calculations for meson valence PDFs.
%
%
\subsection{Exact calculation}
\label{ssec:LF}
\begin{figure}
\includegraphics[scale=0.5]{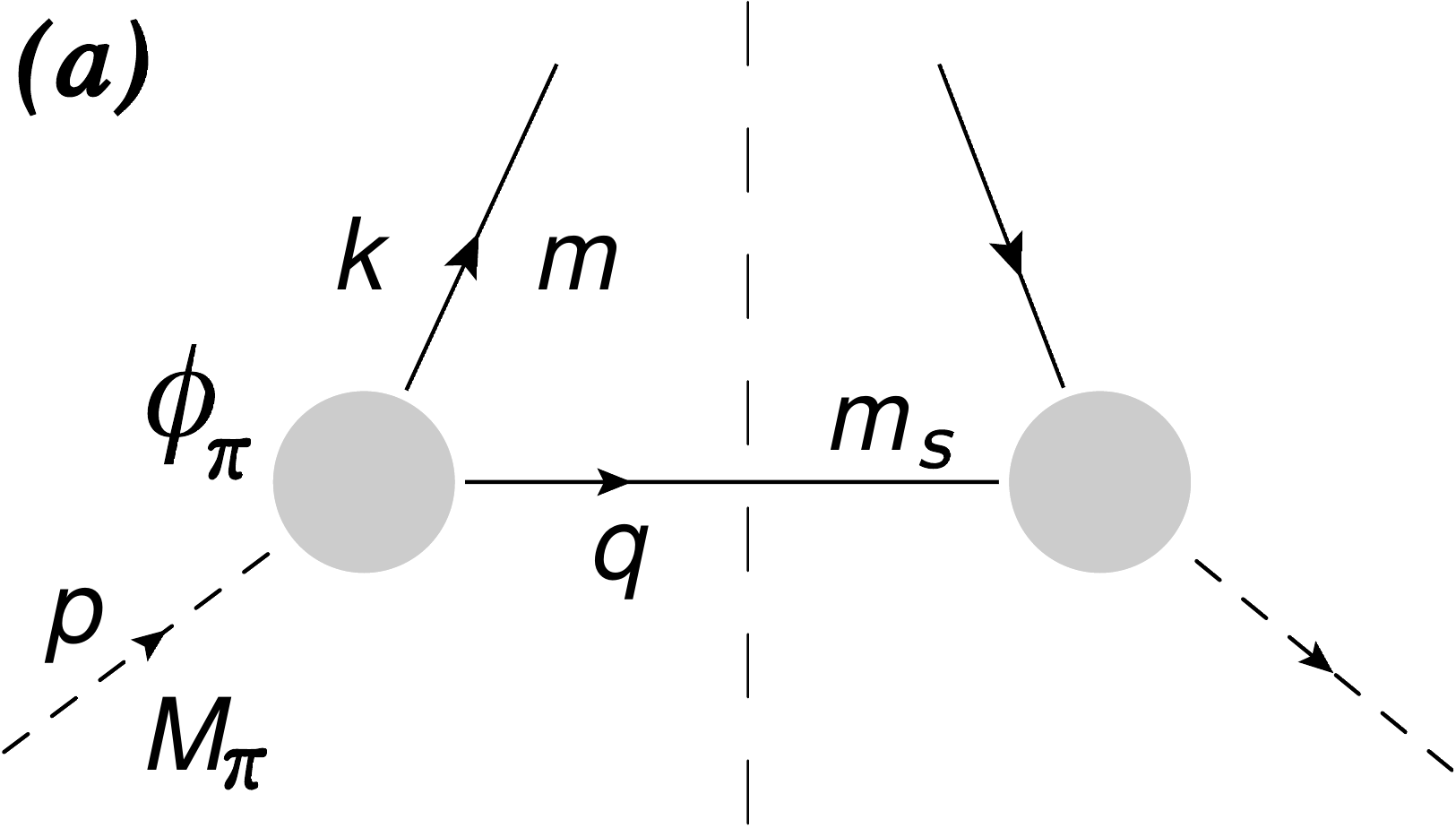} \ \ \ \ \
\includegraphics[scale=0.5]{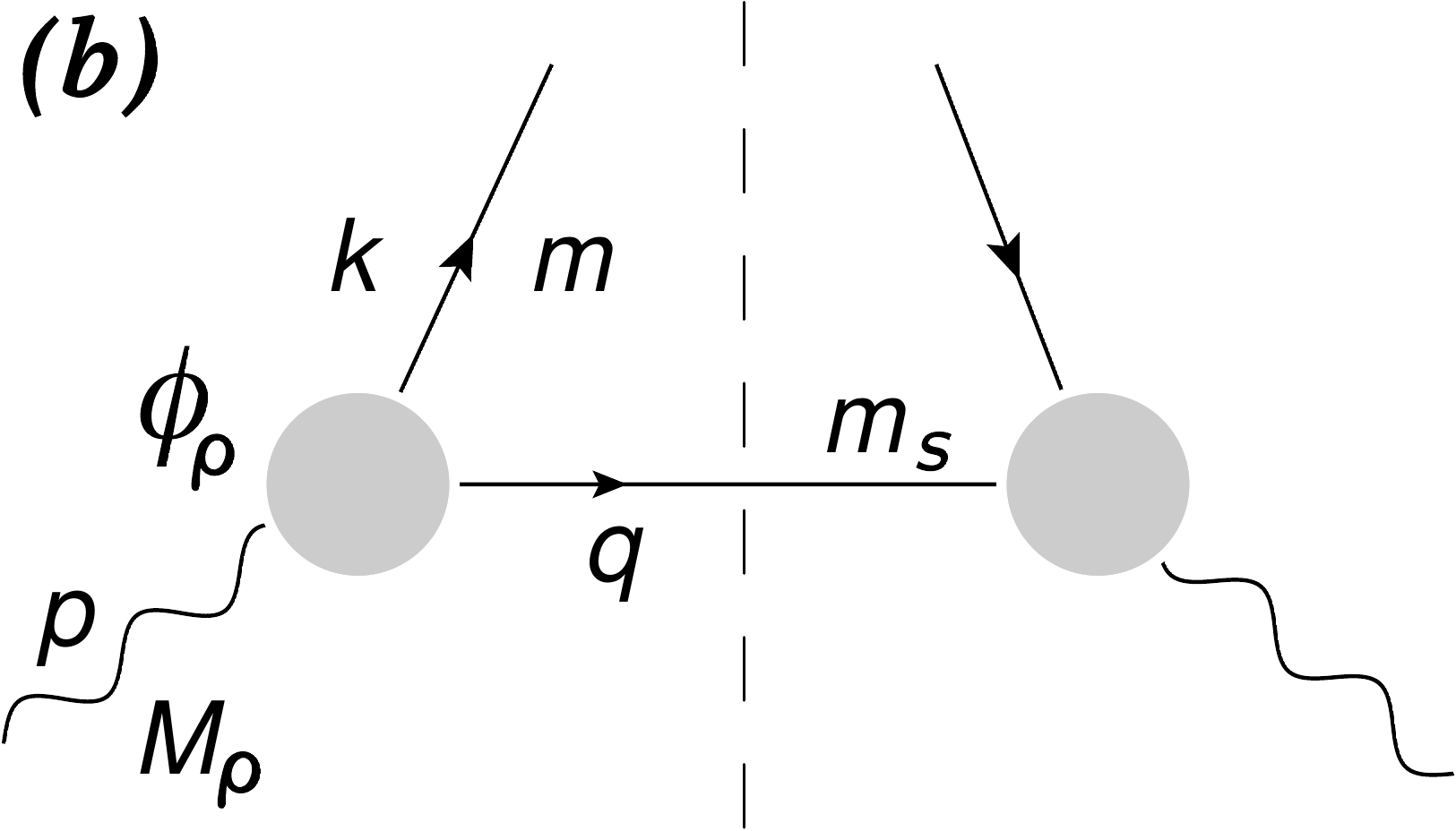}
\caption{(Color online) ({\bf a}) The diagram contributing to the leading-twist valence
	distribution of the pion in the relativistic constituent quark model following
	from the interactions defined in the interactions of Eq.~(\ref{eq:Lpion}). ({\bf b})
	The corresponding graph for the spin-$1$ $\rho$ meson following the Pauli interaction
	of Eq.~(\ref{eq:Lrho}). In the model, the leading twist cut corresponds to a photon
	resolving a struck quark of momentum $k$ emerging from an initial-state meson with
	external momentum $p$ with a spectator quark carrying momentum $q = p - k$. The meson-quark
	vertex is regulated by an appropriate choice of phenomenological function $\phi_M(k^2)$ as
	defined in Eq.~(\ref{eq:phi}). 
}
\label{fig:fig1}
\end{figure}

\paragraph{Pion LF distribution.}
For the purpose of this exploratory analysis, I use a simple pseudoscalar interaction for the $\pi$-quark coupling
according to the Lagrangian
\begin{equation}
	\mathcal{L}_{\pi q s}\ = i N^{1/2}_\pi\ \overline{\psi}_q \gamma_5 \varphi_\pi \psi_s\,  +\ \mathrm{h.\,c.}\ ,
\label{eq:Lpion}
\end{equation}
in which $\psi_{q,\,s}$ denote the intermediate quark fields depicted in Fig.~\ref{fig:fig1}({\bf a}) and $\varphi_\pi$ represents
the pseudoscalar $\pi$ field. The underlying process depicted in Fig.~\ref{fig:fig1} entails the short-lived dissociation of an initial-state
meson of mass $M_{\pi,\,\rho}$ carrying momentum $p$ into an intermediate state involving its constituent quark degrees of freedom ---
an interacting quark of mass $m$ with momentum $k$ and a recoiling spectator quark of mass $m_s$ and momentum $q=p-k$ by conservation.
Taking this as the basis for the meson model, for the exact LF valence quark distribution in the $\pi$, following
the interaction defined by Eq.~(\ref{eq:Lpion}) and standard Feynman rules for Fig.~\ref{fig:fig1}({\bf a}), I write down the
expression
\begin{align}
	q^{\mathrm{LF}}_\pi (x)\ =\ {N_\pi \over 2\, (2\pi)^4} \int dk^+ dk^- d^2k_\perp\ &\left({ 1 \over 2 p^+ }\right)\,
	\delta \left( x - {k^+ \over p^+} \right) \nonumber\\
	&\times\ \mathit{tr} \Big( \gamma_5\, (\ksl + m)\, \gamma^+ (\ksl +m) \gamma_5 (-\qsl + m_s) \Big)\
	2\pi\, \delta \big( q^2 - m^2_s \big) \left[ \phi_\pi(k^2) \over (k^2 - m^2) \right]^2\ ,
\label{eq:piLF}
\end{align}
wherein I have used the standard definition \cite{Kang:2008ey,Collins:2011zzd} of the forward cut of the struck quark line to isolate the
leading-twist (LT) contribution, {\it i.e.},
\begin{equation}
	\mathrm{LT\ quark\!-\!line\ cut}\ \longrightarrow\ {\gamma^+ \over 2p^+}\, \delta \left( x - {k^+ \over p^+} \right)\ .
\label{eq:cut1}
\end{equation}
The overall normalization $N_\pi$ of the $\pi$ valence quark distribution $q^{\mathrm{LF}}_\pi(x)$ is related to the $\pi$-quark
interaction strength of Eq.~(\ref{eq:Lpion}) and by probability conservation must satisfy the condition
\begin{equation}
	\int_0^1 dx\, q_v(x)\ =\ 1\ ,
\label{eq:norm1}
\end{equation}
so that the overall normalization is automatically determined according to
\begin{equation}
	N_\pi\ =\ 1 \bigg/ \int dx\, dk^2_\perp\, q^{\mathrm{LF}}_\pi(x, k^2_\perp)\ ;
\label{eq:norm2}
\end{equation}
this in turn ensures that Eq.~(\ref{eq:norm1}) is always true by construction.
%

%
The quark-meson interaction leading to Eq.~(\ref{eq:piLF}) is inherently nonperturbative and in phenomenological models requires
the introduction of a vertex factor to suppress ultraviolet divergences at large $k$. Here, I use a simple function of the
struck quark's squared momentum,
\begin{equation}
	\big[ \phi_M(k^2) \big]{}^2\ \equiv\ \Lambda^2_M \Big/ \big( k^2 - \Lambda^2_M \big)\ \hspace{2cm} M = (\pi,\ \rho)\ ,
	\label{eq:phi}
\end{equation}
for which I tune in Sect.~\ref{sec:numerics} the values of the parameters $\Lambda_M$ to reproduce structure function moments for the $\pi$ and $\rho$
mesons that have been computed via lattice techniques \cite{Best:1997qp}. In principle, regulator functions with stronger
dependence on $k^2$ are possible, but I proceed with this analysis using the form in Eq.~(\ref{eq:phi}), given that this
guarantees the $q_v(x) \sim (1-x)$ behavior at $x \to 1$ expected from power-counting logic and suggested by the limited
data on the $\pi$ structure function obtained via the Drell-Yan process \cite{Conway:1989fs}. Also, in general
expressions like Eq.~(\ref{eq:phi}) above, quantities that depend upon the choice of initial state meson ({\it i.e.},
either $\pi$ or $\rho$) are indexed with the subscript $M$ for compactness. 
%

%
%
\begin{figure}
\includegraphics[scale=0.31]{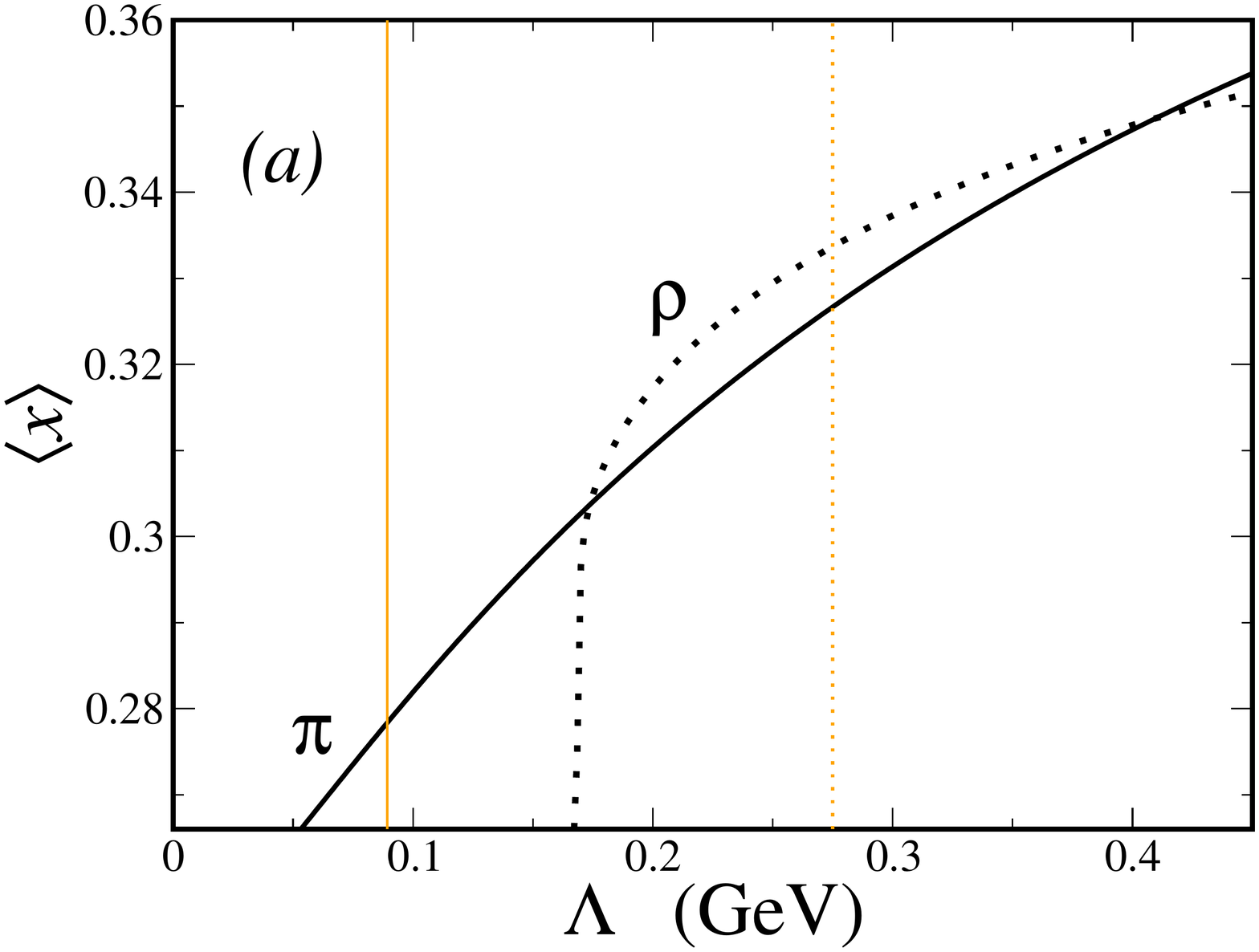} \ \ \ \ \
\raisebox{0cm}{\includegraphics[scale=0.31]{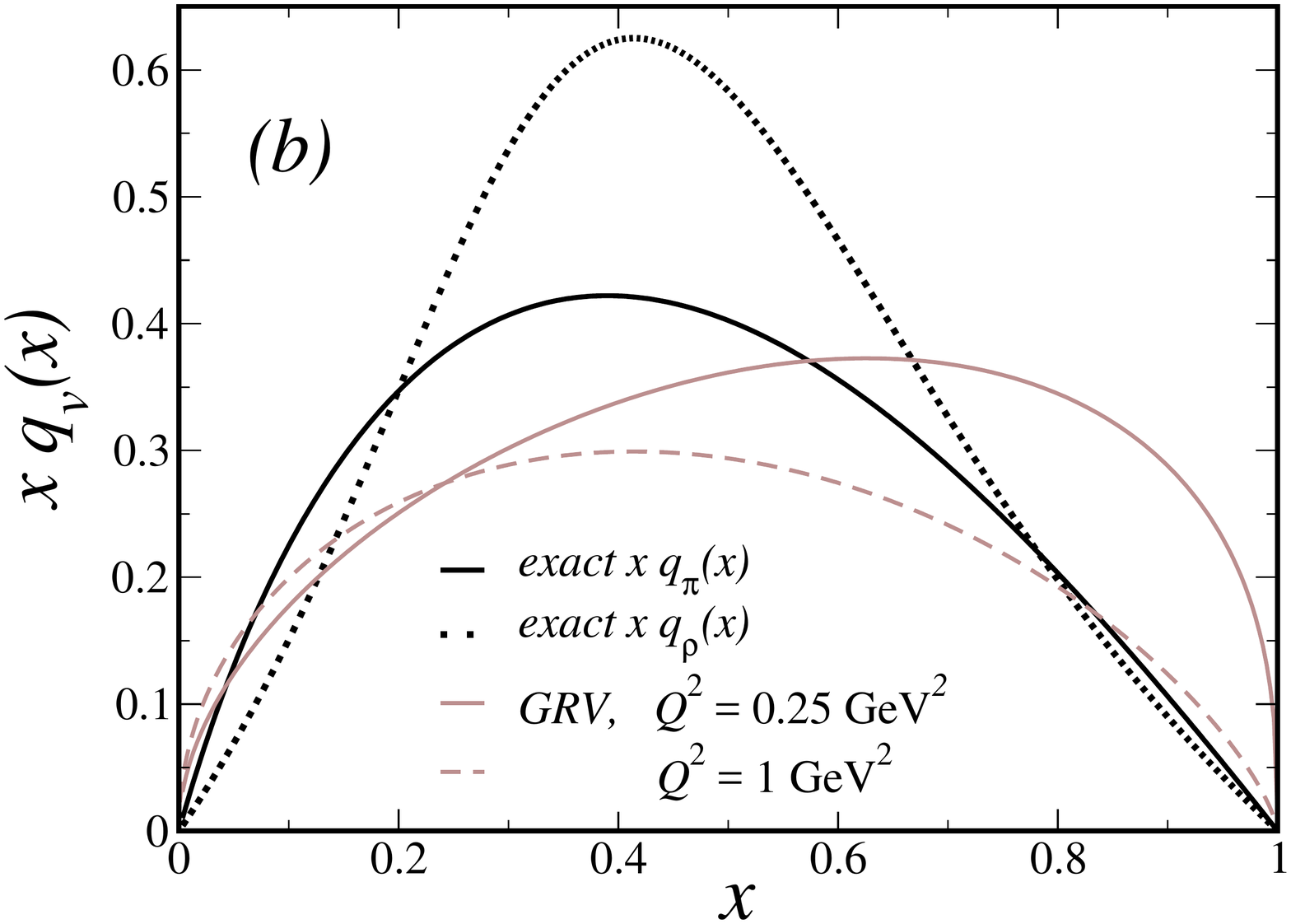}}
\caption{(Color online) ({\bf a}) A plot of the dependence of the $1^{st}$ moments of the mesons' exact LF
	valence distributions $\lan x \ran$ as a function of the UV regulators $\Lambda_M$ appearing in Eq.~(\ref{eq:phi}) for both the
	$\pi$ (solid line) and $\rho$ (dotted line). The associated vertical bars mark the specific values
	to which I tune the model so as to reproduce the lattice-computed moments for these distributions
	found in Ref.~\cite{Best:1997qp}. ({\bf b}) The exact LF valence distributions $xq_v(x)$ corresponding to these choices
	for $\Lambda_M$. The linestyles for the darker curves match those of panel ({\bf a}), while the lighter brown curves
	are the output of the GRV $\pi$ distribution global analysis in Ref.~\cite{Gluck:1991ey} at two low scales of $Q^2$ as
	noted in the legend.
}
\label{fig:fig2}
\end{figure}
Evaluating the expression in Eq.~(\ref{eq:piLF}) according to standard techniques, integrating the $\delta$-function
in $q^2 - m^2_s$ over $k^-$ and using the on-shellness of the struck quark on the light-front, I obtain the {\it exact}
expression for the $\pi$ valence quark PDF,
\begin{align}
\label{eq:piLFfin}
	q^{\mathrm{LF}}_\pi (x)\ &=\ {N_\pi \over 8\pi^2} \int {dk^2_\perp \over x^2 (1-x)}\
	\Big\{ k^2_\perp + \big( x m_s + (1-x)\, m \big)^2 \Big\}
	\left[ \phi_\pi(t_\pi) \over (M^2_\pi - \hat{s}) \right]^2\ \\
	\hat{s}\ &=\ {m^2 + k^2_\perp \over x}\, +\,  {m^2_s + k^2_\perp \over 1-x}\ , \nonumber
\end{align}
where $\hat{s}$ denotes the partonic invariant mass of the internal $2$-quark system and I note that, on the light-front, the
covariant squared momentum of the struck quark after integrating $\int dk^- \delta (q^2 - m^2_s)$ is given by
\begin{equation}
	t_M \equiv k^2_M = k^+ k^-_M\, -\, k^2_\perp = {1 \over 1-x}\, \Big(\! -k^2_\perp - xm^2_s + x(1-x) M^2_M\, \Big)\ .
\label{eq:tLF}
\end{equation}
I note that arriving at Eqs.~(\ref{eq:piLFfin})--(\ref{eq:tLF}) requires integrating over $k^+$ and $k^-$ by means of the $\delta$-functions resulting
from the cut in Fig.~\ref{fig:fig1}; namely, these are
\begin{align}
	\delta \left( x - {k^+ \over p^+} \right)\ &=\ p^+ \delta \left( x p^+ - k^+ \right) \nonumber\\
	\delta \left(q^2 - m^2_s\right)\ &=\ {1 \over (1-x)p^+}\, \delta \left( p^- - k^- - {m^2_s + k^2_\perp \over (1-x) p^+} \right)\ ,
\label{eq:LFdeltas}
\end{align}
where I have assumed integration over the first line in writing the second.
\paragraph{Exact rho meson distribution.}
It is possible to carry out a similar calculation for the $\rho$ meson, for which I take a minimal Lagrangian of the form
\begin{equation}
	\mathcal{L}_{\rho q s}\ = N^{1/2}_\rho\ \overline{\psi}_q\, \gamma_\mu\, \psi_s\, \vartheta^\mu_\rho\ +\ \mathrm{h.\,c.}\ ,
\label{eq:Lrho}
\end{equation}
in which the vector field of the spin-$1$ $\rho$ meson is now denoted by $\vartheta^\mu_\rho$.
Using this form of the interaction and evaluating the corresponding diagram shown in Fig.~\ref{fig:fig1}({\bf b}) I get the exact model expression
for the valence quark distribution of the $\rho$ meson,
\begin{align}
	q^{\mathrm{LF}}_\rho (x)\ &=\ {N_\rho \over 2\, (2\pi)^4} \int dk^+ dk^- d^2k_\perp\ \left({ 1 \over 2 p^+ }\right)\,
	\delta \left( x - {k^+ \over p^+} \right) \nonumber\\
	&\times\, \left( g^{\mu\nu} - {1 \over M^2_\rho}\, p^\mu p^\nu \right)\
	\mathit{tr} \Big( \gamma_\mu\, (\ksl + m)\, \gamma^+ (\ksl +m)\, \gamma_\nu (-\qsl + m_s) \Big)\
	2\pi\, \delta \big( q^2 - m^2_s \big) \left[ \phi_\rho(k^2) \over (k^2 - m^2) \right]^2\ .
\label{eq:rhoLF}
\end{align}
In this case, the higher spin of the $\rho$ generates a more convoluted momentum dependence from the trace algebra in the numerator of
Eq.~(\ref{eq:rhoLF}) above, ultimately leading to
\begin{align}
	q^{\mathrm{LF}}_\rho (x)\ =\ {N_\rho \over 8\pi^2} \int {dk^2_\perp \over x^2 (1-x)^2}\
	\Big\{ k^2_\perp &+ 4x(1-x)m m_s + \big(\,x m_s + (1-x)\, m \big){}^2\ \nonumber\\
	&+\ \left( m^2 + k^2_\perp + x^2\, M^2_\rho \right)
	\left( {m^2_s + k^2_\perp \over M^2_\rho} + (1-x)^2 \right) \Big\}
	\left[ \phi_\rho(t_\rho) \over (M^2_\rho - \hat{s}) \right]^2\ .
\label{eq:rhoLFfin}
\end{align}
I plot my numerical determination for this distribution alongside that of the $\pi$ in Fig.~\ref{fig:fig2}
and describe the associated numerical method in Sect.~\ref{sec:numerics} after describing the corresponding
LaMET calculation below.
%
%
%
\subsection{LaMET approach}
\label{ssec:LaMET}
In the LaMET formalism the cut in the struck quark lines of Fig.~\ref{fig:fig1} given by Eq.~(\ref{eq:cut1}) for
the exact calculation must be replaced by the modified leading-twist operator insertion
\begin{equation}
	{\gamma^+ \over 2p^+}\, \delta \left( x - {k^+ \over p^+} \right)\
	\longrightarrow\ {\gamma^z \over 2p_z}\, \delta \left( x - {k_z \over p_z} \right)
\label{eq:cut}
\end{equation}
such that in the present constituent quark model Eq.~(\ref{eq:piLF}) becomes for the $\pi$ quasi-PDF
\begin{align}
	\widetilde{q}_\pi (x,p_z)\ =\ {N_\pi \over (2\pi)^4} \int dk^0 &dk_z\, d^2k_\perp\ \left({ 1 \over 2 p_z }\right)\,
	\delta \left( x - {k_z \over p_z} \right) \nonumber\\
	&\times\ \mathit{tr} \Big( \gamma_5\, (\ksl + m)\, \gamma^z (\ksl +m) \gamma_5 (-\qsl + m_s) \Big)\
	2\pi\, \delta \big( q^2 - m^2_s \big) \left[ \phi_\pi(k^2) \over (k^2 - m^2) \right]^2\ .
\label{eq:quasipi}
\end{align}
For the quasi-PDF calculation at finite $p_z$ I require explicit expressions in terms of $x$ and $k^2_\perp$ that result
from integrating the $\delta$-function of Eq.~(\ref{eq:quasipi}), $\int dk^0 \delta ( q^2 - m^2_s )$, according
to the Cutkosky procedure; this again has the effect of rendering the kinematics as appropriate for an on-shell spectator
particle. As with the LF calculation before, doing so results in the analogue of Eq.~(\ref{eq:LFdeltas}), but for the
LaMET quasi-distributions; I use
\begin{align}
	\delta \left( x - {k_z \over p_z} \right)\ &=\ p_z\, \delta \left( x p_z - k_z \right) \nonumber\\
	\delta \left(q^2 - m^2_s\right)\ &=\ {1 \over 2 \left( p^0 - k^0 \right) }\, \delta \left( p^0 - k^0
	- \sqrt{m^2_s + k^2_\perp + (1-x)^2 p^2_z} \right)\ .
\label{eq:LMETdeltas}
\end{align}
I denote quantities subsequent to the integrations over $k^0$ and $k_z$ in Eq.~(\ref{eq:quasipi}) using
Eq.~(\ref{eq:LMETdeltas}) as $\tilde{p}_M$, {\it etc}. In particular, several scalar products are needed:
\begin{align}
	\tilde{p}_M \cdot \tilde{k}_M\ &=\ \tilde{p}{}^{\,0}_M\, \tilde{k}{}^0_M - x p^2_z \nonumber \\
	\tilde{p}_M \cdot \tilde{q}\ &=\ \tilde{p}{}^{\,0}_M\, \tilde{q}{}^{\,0} - (1-x) p^2_z \nonumber \\
	\tilde{q} \cdot \tilde{k}_M\ &=\ \tilde{q}{}^{\,0}\, \tilde{k}{}^0_M + k^2_\perp - x (1-x) p^2_z\ ,
\label{eq:LMETdots}
\end{align}
for which the relevant energies $\tilde{p}{}^{\,0}_M$, {\it etc}., are
\begin{align}
	\tilde{p}{}^{\,0}_M\ &\equiv\ p_z\, \mu_M  \nonumber\\
	\tilde{q}{}^{\,0}\ &\equiv\ (1-x) p_z \mu_s \nonumber\\
	\tilde{k}{}^0_M\ &\equiv\ \sqrt{ M^2_M + m^2_s + k^2_\perp + \big(1+[1-x]\,\big){}^2\, p^2_z - 2(1-x) \big(1 - \mu_M\,\mu_s \big) p^2_z } \nonumber\\
	\mu_M\ &\equiv\ \sqrt{ 1 + {M^2_M \over p^2_z} } \hspace{1cm} \mu_s\ \equiv\ \sqrt{ 1 + {m^2_s + k^2_\perp \over (1-x)^2 p^2_z} }\ ,
\end{align}
and similarly,
\begin{align}
	\tilde{k}^2_M\ &\equiv\ \big(\tilde{k}^0_M\big){}^2 - k^2_\perp - x^2 p^2_z\ =\ M^2_M + m^2_s + 2\, (1-x) \big( 1 - \mu_M\, \mu_s \big)\, p^2_z\ .
\label{eq:LMETk2}
\end{align}

Evaluating the trace and integrating over $k^0$ and $k_z$ in Eq.~(\ref{eq:quasipi}) using the $\delta$-function expressions of Eq.~(\ref{eq:LMETdeltas}) leads
to the $\pi$ quasi-distribution in terms of the quantities defined in Eqs.~(\ref{eq:LMETdots})--(\ref{eq:LMETk2}):
\begin{align}
	\widetilde{q}_\pi (x,p_z)\ =\ {N_\pi \over 4 \pi^2} \int {dk^2_\perp \over 2(1-x)\mu_s}\ 
	&\left\{ 2x \left(  m m_s + (\tilde{q}\cdot\tilde{k}_\pi) \right)
	+ ( m^2 - \tilde{k}^2_\pi ) \left( 1-x \right) \right\} \nonumber\\
	&\hspace{2cm}\times \left[ \phi_\pi(\tilde{k}^2_\pi) \over (M^2_\pi + m^2_s - m^2 + 2(1-x)\big( 1 - \mu_\pi\,\mu_s \big) \right]^2\ ;
\label{eq:qpi_fin}
\end{align}
this expression may then be evaluated numerically at the same values of $m$, $m_s$, $\Lambda_\pi$, and $N_\pi$ found for the exact LF
calculation of Eq.~(\ref{eq:piLFfin}) by integrating over $k^2_\perp$. Moreover, I examine $\widetilde{q}_\pi (x,p_z)$ at finite values of
$p_z$ in Sect.~\ref{sec:numerics} below and sketch the extent to which the high $x$ valence region of $q_\pi(x)$ may be accessed with the
LaMET method.
The same formalism can be used for the $\rho$ to obtain an analogous $p_z$ dependent expression for the valence quasi-PDF.
Using the LaMET treatment of the cut quark line in Eq.~(\ref{eq:cut}) and writing the appropriate expression akin to
Eq.~(\ref{eq:rhoLF}) I get
\begin{align}
	\widetilde{q}_\rho (x)\ &=\ {N_\rho \over (2\pi)^4} \int dk^0 dk_z d^2k_\perp\ \left({ 1 \over 2 p_z }\right)\,
	\delta \left( x - {k_z \over p_z} \right) \nonumber\\
	&\times\ \left( g^{\mu\nu} - {1 \over M^2_\rho}\, p^\mu p^\nu \right)\
	\mathit{tr} \Big( \gamma_\mu\, (\ksl + m)\, \gamma^z (\ksl +m)\, \gamma_\nu (-\qsl + m_s) \Big)\
	2\pi\, \delta \big( q^2 - m^2_s \big) \left[ \phi_\rho(k^2) \over (k^2 - m^2) \right]^2\ ,
\label{eq:qrho}
\end{align}
which, like the $\pi$ quasi-distribution, closely mirrors its exact LF analogue, aside from its equal-time form and space-like
operator structure. Evaluating this last expression as before at the relevant finite-$p_z$ kinematics again yields the desired formula
in terms of the inner products of Eqs.~(\ref{eq:LMETdots})--(\ref{eq:LMETk2}):
\begin{align}
	\widetilde{q}_\rho (x,p_z)\ &=\ {N_\rho \over 4 \pi^2} \int {dk^2_\perp \over (1-x)\mu_s}\
	\left\{ x \left( {2\over M^2_\rho} (\tilde{p}_\rho\cdot\tilde{q})(\tilde{p}_\rho\cdot\tilde{k}_\rho) + 3m m_s 
	+ (\tilde{q}\cdot\tilde{k}_\rho) \right)
	+ {1\over 2} (m^2 - \tilde{k}^2_\rho ) \left( {2 (\tilde{p}_\rho\cdot\tilde{q}) \over M^2_\rho} + (1-x) \right) \right\} \nonumber\\
	&\times \left[ \phi_\rho(\tilde{k}^2_\rho) \over (M^2_\rho + m^2_s - m^2 + 2(1-x)\big( 1 - \mu_\rho\,\mu_s \big) \right]^2\ .
\label{eq:qrho_fin}
\end{align}
This latter expression can also be computed for plausible values in the model parameter space, and I consider the side-by-side
comparison of the LaMET {\it vs.} exact LF calculations of $q_\rho(x)$ along with $q_\pi(x)$ in Sect.~\ref{sec:numerics} below.
%
%
\section{Numerical results}
\label{sec:numerics}
In Sect.~\ref{sec:formal} I used the light-front and LaMET formalisms to derive the (quasi-)PDFs for the $\pi$ and $\rho$
valence quark content; this was done under the auspices of an effective theory that couples the mesons to their constituent quark
content according to the minimal Lagrangians of Eqs.~(\ref{eq:Lpion}) and (\ref{eq:Lrho}). With this formalism established,
I proceed in the present Section to fix the adjustable parameters that enter the constituent quark model, facilitating an analysis of the size and
$x$ dependence of the finite $M^2 \big/ p^2_z$ corrections of Eq.~(\ref{eq:LMETform}) for both the $\pi$ and $\rho$
distributions.
Taking this as the primary goal of the present analysis --- to study the ranges of validity of LaMET approximations to the exact meson
PDFs in a {\it typical} model --- it is sufficient to make plausible, physically-motivated
selections for the model parameters (mainly, the constituent quark masses $m$ and $m_s$ that enter the meson distributions derived in
Sect.~\ref{sec:formal}). With these fixed, it is then possible to tune the remaining degrees of freedom [the UV cutoffs $\Lambda_M$
in Eq.~(\ref{eq:phi})] to reproduce information determined in lattice QCD calculations (the distribution moments), thereby completely
determining the constituent quark model and providing a basis for comparing the LaMET and LF meson distributions in the valence region.
For both the $\pi$ and $\rho$ distributions, I fix hadronic masses to their pole values $M_\pi = 0.139$ GeV and $M_\rho = 0.77$ GeV.
In the case of the intermediate states, I take for the constituent masses of the $\pi$ valence quarks the static values
$m = m_s = M/3 \approx 0.33$ GeV, where $M$ is the proton mass. For the heavier $\rho$ meson, however, this choice would introduce
numerical instabilities due to the presence of the quark denominators in Eq.~(\ref{eq:rhoLF}), which produce potentially singular overall factors of
$[M_\rho - \hat{s}]^{-2}$ in the final expression for $q^{\mathrm{LF}}_\rho(x)$ in Eq.~(\ref{eq:rhoLFfin}). Essentially, the model in this
circumstance is not sufficiently confining to prevent the decay of the $\rho$ into its lighter-mass partonic constituents. A simple repairative
scheme to address this issue is the effective mass prescription \cite{Pumplin:2005yf,Hobbs:2013bia}, which involves the selection of a larger
mass for the bound constituent quark; this is on the logic that the nonperturbative interactions necessary for confinement in this case dynamically
generate a larger effective mass for the constituent quark in the $\rho$. For the current analysis, I proceed with this assumption,
taking the effective quark masses in the $\rho$ to be a conservative value, $m = m_s = m^{\mathit{eff}} \equiv 0.6$ GeV.
%

%
%
\begin{figure}
\includegraphics[scale=0.31]{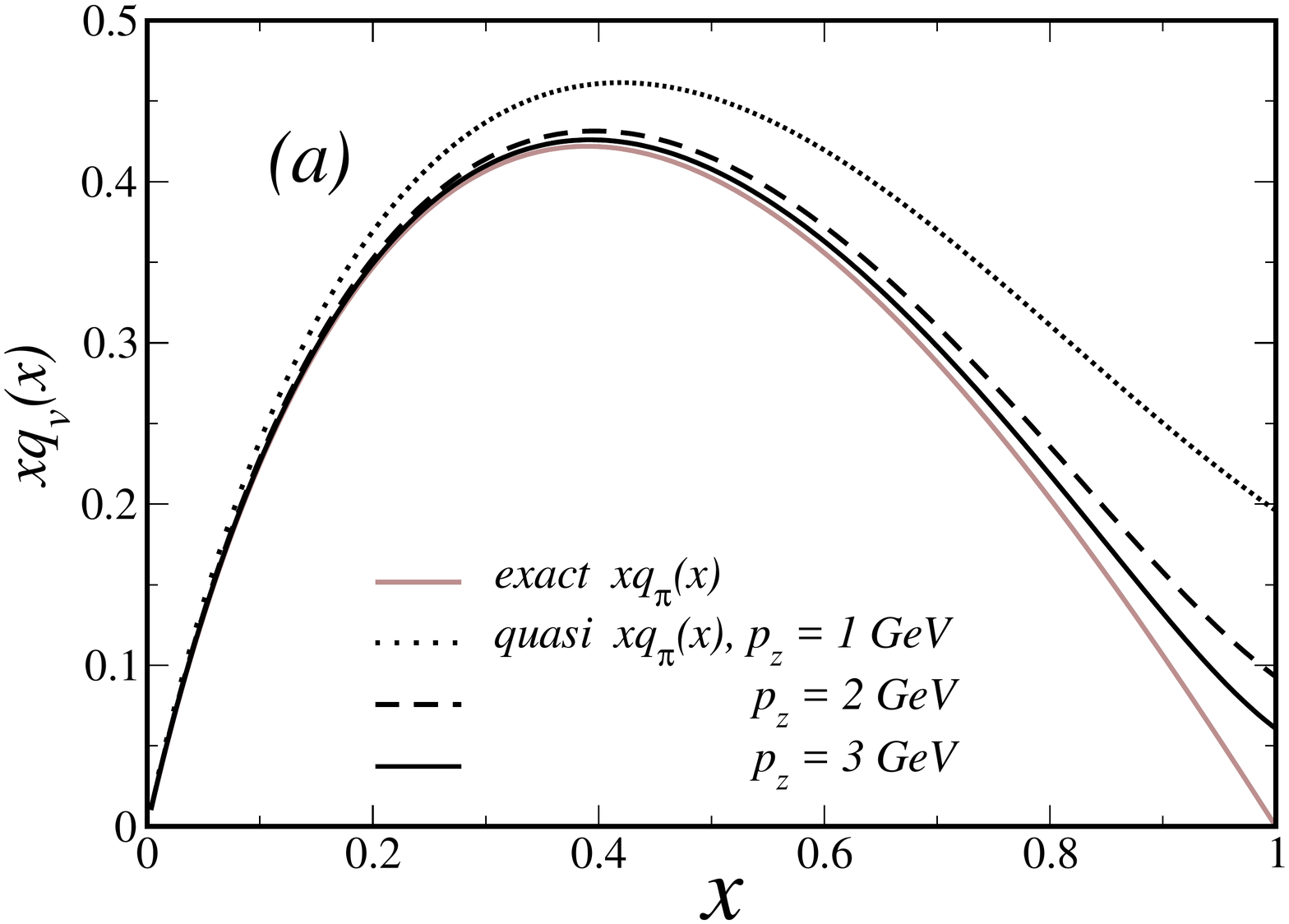} \ \ \ \ \
\raisebox{0cm}{\includegraphics[scale=0.31]{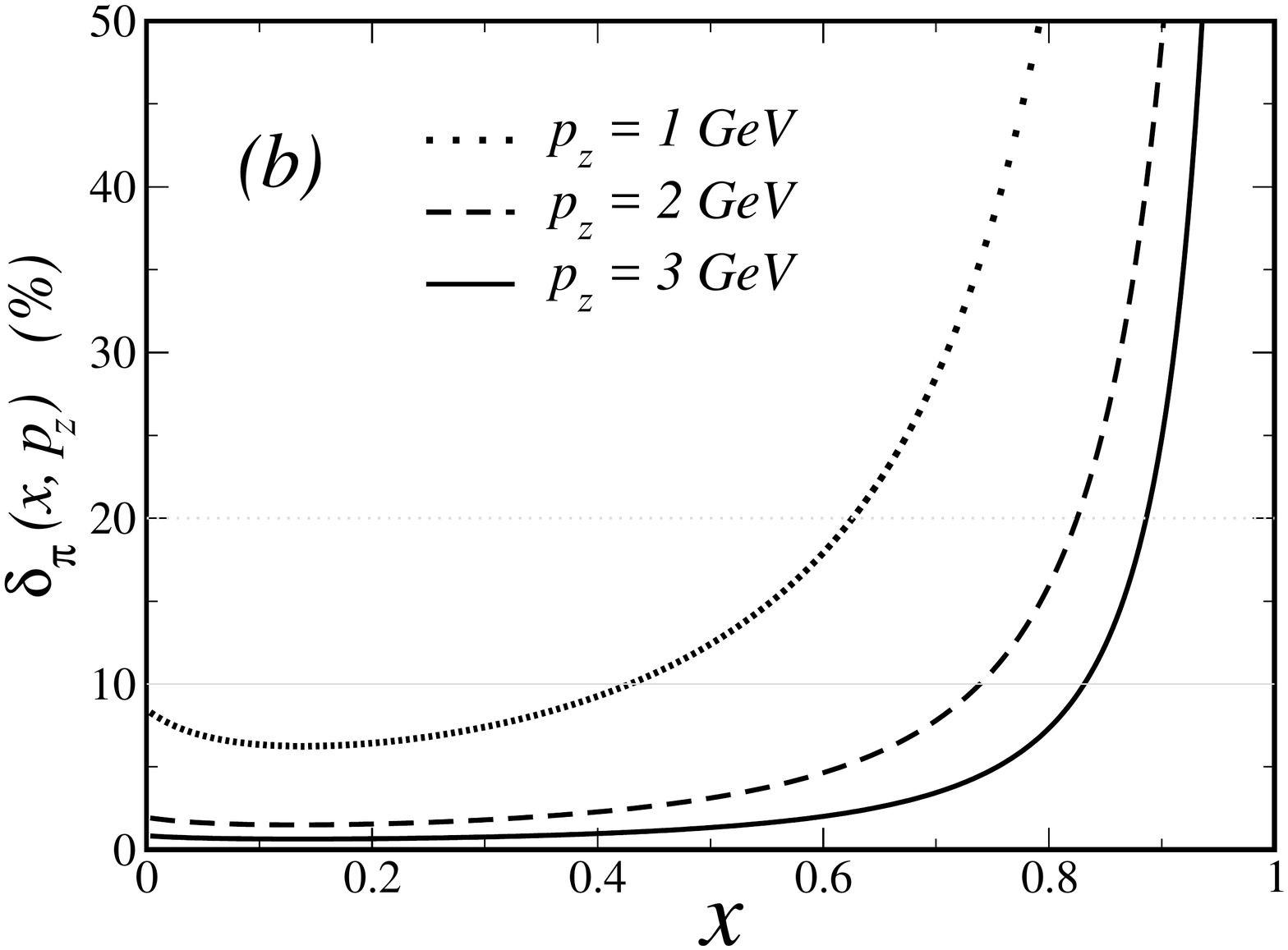}}
\caption{(Color online) ({\bf a}) A plot of the exact $\pi$ distribution
	function $q^{\mathrm{LF}}_\pi(x)$ (lighter brown, solid) alongside the LaMET quasi-PDF
	$\widetilde{q}_\pi(x,p_z)$ (heavy black curves) for several choices of $p_z = 1$ GeV
	(dotted), $2$ GeV (dashed), and $3$ GeV (solid). In the limit of arbitrarily
	large boosts, $p_z \to \infty$, the black curves match onto the brown. ({\bf b}) To
	illustrate the $x$ dependent difference between the LaMET approximation
	and exact valence PDF, I plot the discrepancy parameter $\delta_\pi(x,p_z)$ 
	defined in Eq.~(\ref{eq:disc}), keeping the linestyles here as they were in
	panel ({\bf a}).
}
\label{fig:fig3}
\end{figure}
With the mass parameters of the LF model thus fixed, and the overall normalization $N_M$ determined by Eq.~(\ref{eq:norm2}), it remains
to set the shape parameters $\Lambda_M$ that enter the relativistic vertex function in Eq.~(\ref{eq:phi}). To place the exact LF model
in concordance with extant lattice information on the $\pi$ and $\rho$ structure functions, I tune the cutoff parameters such that
the $1^{st}$ moment of the mesons' valence distributions,
\begin{equation}
	\lan x \ran_M\ =\ \int_0^1 dx\, x\!\,  q^{\mathrm{LF}}_M (x)\ ,
\label{eq:xmom}
\end{equation}
agrees with the central values predicted by the lattice calculation reported in Ref.~\cite{Best:1997qp}. As might be expected on the basis of
Eqs.~(\ref{eq:piLFfin}) and (\ref{eq:rhoLFfin}), the distributions $q^{\mathrm{LF}}_M(x)$ as well as their $1^{st}$ moments $\lan x \ran_M$
depend upon $\Lambda_M$ in a nonlinear fashion as shown in Fig.~\ref{fig:fig2}({\bf a}).
Performing the necessary tuning of $\Lambda_M$ for the $\pi$ and $\rho$, respectively, therefore leads to the rather different values
\begin{align}
	\lan x \ran_\pi\  &=\  0.279(83) \hspace{0.75cm} \longrightarrow \hspace{0.75cm}   \Lambda_\pi\ =\ 0.0892\, \mathrm{GeV} \nonumber \\
	\lan x \ran_\rho\ &=\  0.334(21) \hspace{0.75cm} \longrightarrow \hspace{0.75cm}   \Lambda_\rho\ \hspace{0.04cm} =\ 0.2750\,  \mathrm{GeV}\ ,
\label{eq:Lambs}
\end{align}
which correspond to the vertical bars of Fig.~\ref{fig:fig2}({\bf a}). Setting the UV cutoffs according to the values of Eq.~(\ref{eq:Lambs})
and using the mass parameter choices mentioned above fully determines the exact LF distributions $q^{\mathrm{LF}}_{\pi,\, \rho}(x)$ of
Eqs.~(\ref{eq:piLFfin}) and (\ref{eq:rhoLFfin}), which I plot in Fig.~\ref{fig:fig2}({\bf b}). For comparison, the LF distributions of the
$\pi$ and $\rho$ meson are plotted alongside the GRV extraction \cite{Gluck:1991ey} at two lower scales, $Q^2 = 0.25, 1$ GeV$^2$, of the $\pi$
valence quark distribution from the global analysis of the E615 Drell-Yan data \cite{Conway:1989fs}. This comparison illustrates that a minimal
model with few degrees of freedom is capable of falling well within the range of purely data-driven determinations.
%

%
%
\begin{figure}
\includegraphics[scale=0.31]{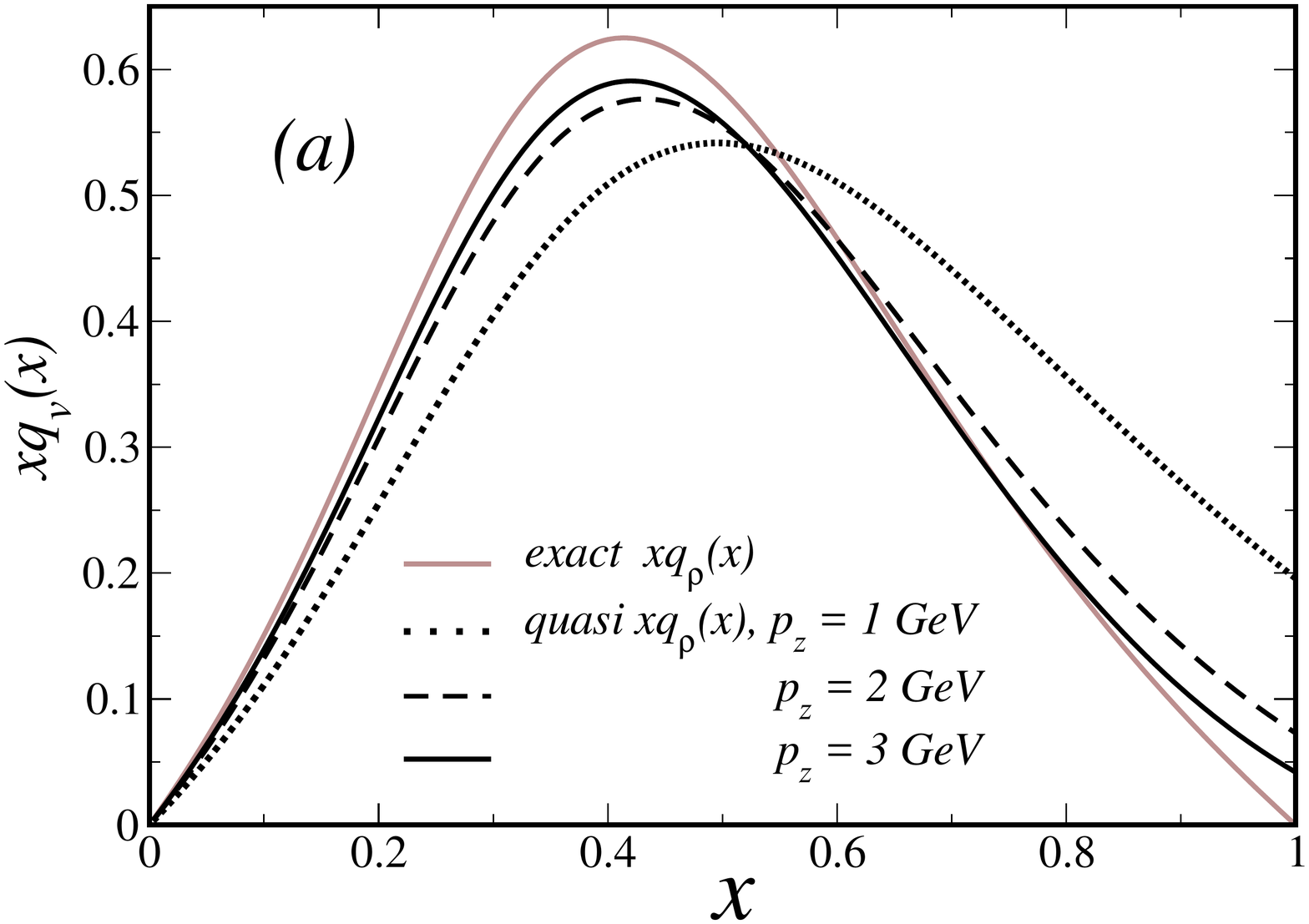} \ \ \ \ \
\raisebox{0cm}{\includegraphics[scale=0.31]{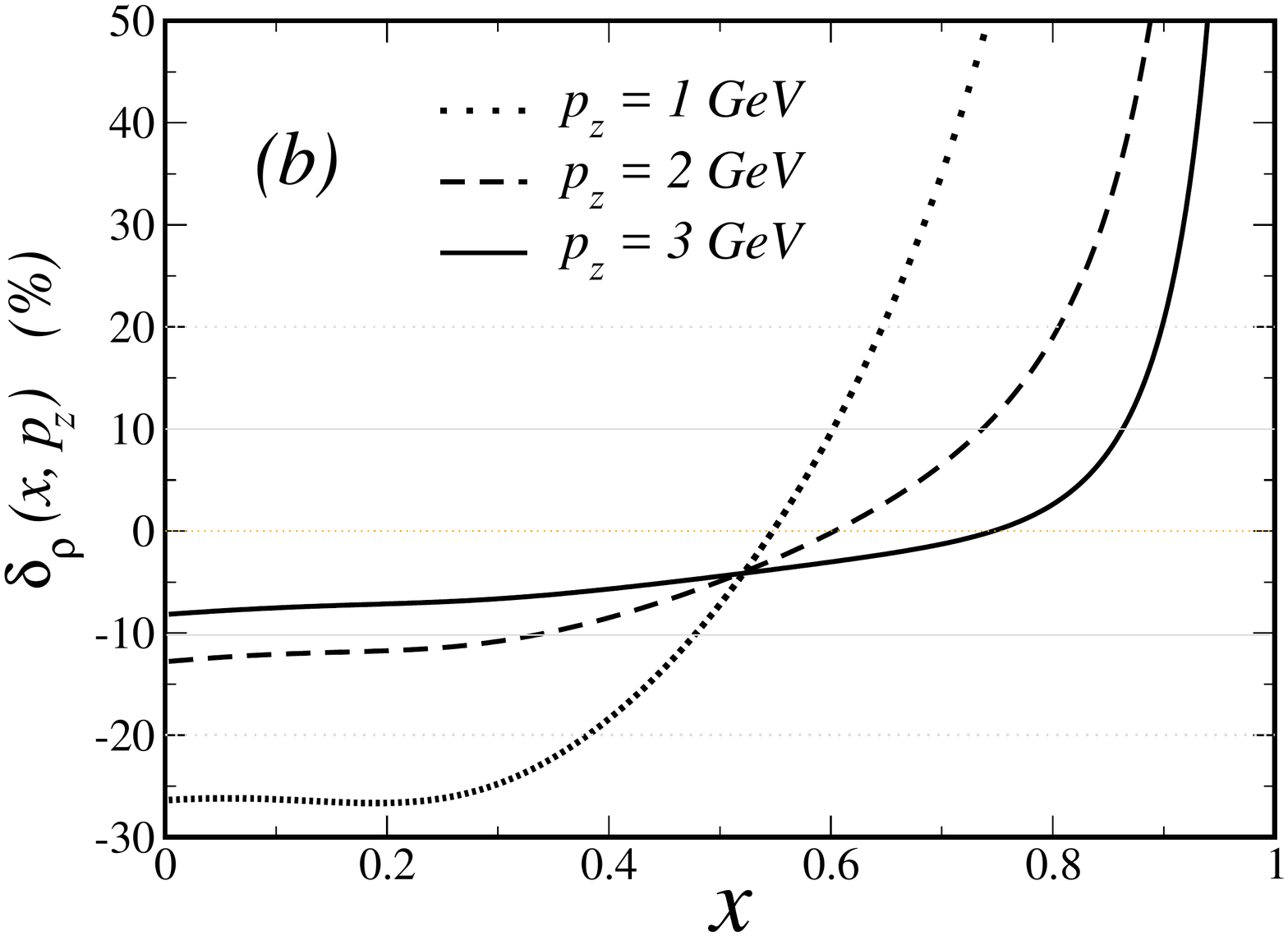}}
	\caption{(Color online) Analogous to Fig.~\ref{fig:fig3} for the $\pi$, panel 
	({\bf a}) shows the exact and quasi-PDFs for the $\rho$ meson's valence structure, while ({\bf b})
	gives the associated discrepancy parameter $\delta_\rho(x, p_z)$.
}
\label{fig:fig4}
\end{figure}
With the exact LF model for the $\pi$ and $\rho$ valence distributions thus in place, the finite-$p_z$ corrections to the LaMET
quasi-distributions --- the main result of this analysis --- may be computed directly. Doing so involves evaluating Eqs.~(\ref{eq:qpi_fin})
and (\ref{eq:qrho_fin}) of Sect.~\ref{sec:formal} with the same parameters detailed above for the exact LF calculation, keeping the overall
normalization $N_M$ fixed as well; I choose several finite values of $p_z$ at which to compute, taking $p_z = \{1,\, 2,\, 3\ \mathrm{GeV}\}$ as
representative choices for the hadronic boost that roughly corresponds to the intermediate range of previous calculations \cite{Gamberg:2014zwa}
chosen as being accessible to lattice computations. This procedure allows the direct numerical verification of the relation
\begin{equation}
	\lim_{p_z \to \infty}\, \widetilde{q}_M(x, p_z)\ =\ q^{\mathrm{LF}}_M(x)\ ,
\label{eq:LMETlim}
\end{equation}
which represents the fundamental result of LaMET applied to mesonic structure. It is possible to make this verification analytically by
considering the behavior of Eqs.~(\ref{eq:qpi_fin}) and (\ref{eq:qrho_fin}) in the limit of infinite momentum $p_z$, but the detailed breaking of the
equivalence in Eq.~(\ref{eq:LMETlim}) at finite $p_z$ is highly nontrivial, and it is for such lower hadronic momenta that lattice calculations
are carried out in practice.
To make this behavior explicit, I plot the $\pi$ quasi-distributions $x\widetilde{q}_\pi(x,p_z)$ (heavy black curves) next to their exact LF
counterpart $xq^{\mathrm{LF}}_\pi(x)$ (lighter brown line) in Fig.~\ref{fig:fig3}({\bf a}), which shows the the $\pi$ valence quasi-PDFs steadily
approaching the LF result as $p_z$ is increased from $1$ GeV (dotted line), to $2$ GeV (dashed), to $3$ GeV (solid), at which the point the $x$
dependence of the quasi-PDF very closely mirrors that of the exact distribution --- deviating significantly only for $x \gtrsim 0.8$.
To gauge the numerical size of these finite $p_z$ corrections in $\widetilde{q}(x,p_z)$, I find it useful to construct and plot a discrepancy
parameter, which I define as
\begin{equation}
	\delta_M (x, p_z)\ \equiv\ { \widetilde{q}_M(x,p_z) \over q^{\mathrm{LF}}_M(x) }\, -\, 1\ ;
\label{eq:disc}
\end{equation}
this object quantifies as a fractional percentage the size and direction of the deviation of the LaMET meson quasi-PDF from the exact LF
result. For the $\pi$ distributions, I therefore pair Fig.~\ref{fig:fig3}({\bf a}) with a plot of $\delta_\pi (x, p_z)$ for the same
collection of $p_z$ values in Fig.~\ref{fig:fig3}({\bf b}), keeping also the same correspondence between linestyles and $p_z$. A particularly
remarkable feature of these plots is the fact that $x\widetilde{q}_\pi(x,p_z)$ approximates $xq^{\mathrm{LF}}_\pi(x)$ with high
fidelity for $x \lesssim 0.4$ at even the smallest $p_z = 1$ GeV. The nonlinearity of the $p_z$ corrections --- consistent with the
fact that they appear as sub-leading effects $\sim\! 1/p^2_z$ in the formal definition of the quasi-PDF in Eq.~(\ref{eq:LMETform})
--- is also strongly apparent in Fig.~\ref{fig:fig3}; the pronounced gap between the size of $\delta_\pi (x, p_z = 1\, \mathrm{GeV})$
and $\delta_\pi (x, p_z = 2\, \mathrm{GeV})$ in Fig.~\ref{fig:fig3}({\bf b}) suggests that incremental improvements in the reach
in $p_z$ for lattice calculations of $\widetilde{q}_\pi(x,p_z)$ would yield sharply increasing returns in affording precise access
to the high $x$ region of the $\pi$ structure function. For example, with a $p_z$ boost of $2$ GeV, Fig.~\ref{fig:fig3} suggests
that the quasi-PDF approximates the exact calculation to better than $10\%$ for $x \lesssim 0.7$, while affording $\sim\! 20\%$
precision in the high $x$ region until just beyond $x = 0.8$.
The analogous calculation for the $\rho$ meson quasi-PDF is given in Fig.~\ref{fig:fig4}({\bf a}) and \ref{fig:fig4}({\bf b}), which like the panels of
Fig.~\ref{fig:fig3} for the $\pi$ distributions compares the finite-$p_z$ LaMET approximation $x\widetilde{q}_\rho(x,p_z)$ of Eq.~(\ref{eq:qrho_fin})
to the exact result $xq^{\mathrm{LF}}_\rho(x)$ of Eq.~(\ref{eq:rhoLFfin}). Here again, I plot in panel ({\bf a}) the $\rho$ meson quasi-PDFs themselves 
next to $xq^{\mathrm{LF}}_\rho(x)$, while panel ({\bf b}) gives the associated deviations as per $\delta_\rho (x, p_z)$. Noting that performing the
calculation of $\rho$ meson calculations according to Eqs.~(\ref{eq:rhoLFfin}) and (\ref{eq:qrho_fin}) required the use of a large effective mass for
constituent quarks, the different spin structure of the $\rho$ leads to a number of differences in the distribution shapes and behavior with respect to
the $\pi$
calculation. Chief among these are the different distribution curvatures I find (in general more positive for the $\rho$ as opposed to the negative
curvature of the $xq_\pi(x)$ distributions in Fig.~\ref{fig:fig3}), and the fact that the deviations $\delta_\rho(x,p_z)$ of the $\rho$ quasi-PDFs
from the exact result experience a sign-change, unlike $\delta_\pi(x,p_z)$, which remains strictly positive-definite for the parameters considered.
Similarly, the finite-$p_z$ effect in $x\widetilde{q}_\rho(x,p_z)$ for the lowest choice $p_z = 1$ GeV can potentially be of larger fractional size than that
found for the $\pi$ distributions in Fig.~\ref{fig:fig3}; this is especially true at low $x$, although it should be noted that the absolute magnitude of the
discrepancy remains relatively small as can be seen by comparing the curves of Fig.~\ref{fig:fig4}({\bf a}).
As was the case for $x\widetilde{q}_\pi(x,p_z)$, however, these finite-$p_z$ effects rapidly come under control as one moves to larger longitudinal
boosts, and at $p_z = 2$ GeV the fractional deviation is $| \delta_\rho | \lesssim 10\%$ for $x \lesssim 0.75$. This suggests that lattice calculations
ultimately have the potential to achieve similarly precise approximations for the $\rho$ valence distribution as for the $\pi$ via the LaMET method.
%
%
%
%
\section{Conclusions}
\label{sec:conc}
The foregoing represents a preliminary analysis of the collinear quark distributions of light mesons
as might be achieved using the LaMET quasi-PDF formalism. Rather than focusing on the technicalities of lattice gauge theory as
applied to the $\pi$ and $\rho$, I have instead restricted this study to the issue of the $p_z$ corrections which
constitute the main limitation in the application of LaMET techniques to a given hadronic system. As
$2$-quark systems, the $\pi$ and $\rho$ are well-suited to constituent quark models such as the one presented here,
suggesting that these models provide a useful avenue for assessing finite-$p_z$ effects.
The most important conclusion of this analysis is the fact that the corrections arising from finite-$p_z$ effects in the
$\pi$ quasi-distribution $x\widetilde{q}_\pi(x,p_z)$ may be potentially milder than those found in previous analyses
for the nucleon quasi-distribution, including Ref.~\cite{Gamberg:2014zwa}. By this I mean that the size of
the finite-$p_z$ effect in the $\pi$ quasi-PDF at high $x$, {\it e.g.}, $x = 0.7$, is found to be smaller
in the present model calculation ($\delta_\pi \lesssim 10\%$ at $p_z = 2$ GeV) than for the nucleon quasi-PDF
calculation of Ref.~\cite{Gamberg:2014zwa}, which found a factor of $\sim\!\! 2 - 3$ effect for the unpolarized
$xu(x)$ distribution at $p_z = 2$ GeV using a quark-diquark model.
In fact, for $x\widetilde{q}_\pi(x,p_z)$ I predict that the valence PDF should be accessible to lattice calculations
by means of LaMET technology to at least $20\%$ precision for $x \lesssim 0.6$ at hadronic boosts as small as
$p_z = 1$ GeV. At such a small boost, the calculation for the $\rho$ meson quasi-PDF promises to be somewhat more
fraught on the other hand, with the associated $p_z$ corrections correspondingly larger, owing mainly to the
larger $\rho$ mass; extending lattice calculations to higher boosts ({\it e.g.}, $p_z = 2$ GeV) has the potential
to ameliorate these effect, however, and the quasi-PDF discrepancies $\delta_\rho$ of Eq.~(\ref{eq:disc}) begin
to match those of the $\pi$ at these larger momenta.
These observations augur well for the applicability of LaMET-motivated calculations of the light meson PDFs on the
lattice, assuming other technical aspects for carrying out lattice gauge calculations for such light hadrons can
be managed. A lattice computation of the $\pi$ valence quasi-PDF especially has the potential to hugely impact
hadronic structure phenomenology and should be pursued as computational resources become increasingly available.
This study also suggests that a future symbiosis between quark models like the present one and forthcoming lattice
calculations may prove crucial to understanding the results of lattice work --- including the $p_z$ dependence of
computed quasi-distributions and extrapolations thereof to infinite momentum.
%
%
%
%
\section*{Acknowledgements}
I am grateful to Mary Alberg, Silas Beane, Gerald Miller, Fred Olness, and Xilin Zhang for enlightening discussions during the
formative stages of this project.
I also thank Gerald Miller for his critical reading of a draft of the manuscript.
This work was materially supported by the U.S.~Department of Energy Office of Science, Office of Nuclear Physics under
Award No.~DE-FG02-97ER-41014.
%
%
%
\section*{References}
%

\begin{thebibliography}{27}%
\makeatletter
\providecommand \@ifxundefined [1]{%
 \@ifx{#1\undefined}
}%
\providecommand \@ifnum [1]{%
 \ifnum #1\expandafter \@firstoftwo
 \else \expandafter \@secondoftwo
 \fi
}%
\providecommand \@ifx [1]{%
 \ifx #1\expandafter \@firstoftwo
 \else \expandafter \@secondoftwo
 \fi
}%
\providecommand \natexlab [1]{#1}%
\providecommand \enquote  [1]{``#1''}%
\providecommand \bibnamefont  [1]{#1}%
\providecommand \bibfnamefont [1]{#1}%
\providecommand \citenamefont [1]{#1}%
\providecommand \href@noop [0]{\@secondoftwo}%
\providecommand \href [0]{\begingroup \@sanitize@url \@href}%
\providecommand \@href[1]{\@@startlink{#1}\@@href}%
\providecommand \@@href[1]{\endgroup#1\@@endlink}%
\providecommand \@sanitize@url [0]{\catcode `\\12\catcode `\$12\catcode
  `\&12\catcode `\#12\catcode `\^12\catcode `\_12\catcode `\%12\relax}%
\providecommand \@@startlink[1]{}%
\providecommand \@@endlink[0]{}%
\providecommand \url  [0]{\begingroup\@sanitize@url \@url }%
\providecommand \@url [1]{\endgroup\@href {#1}{\urlprefix }}%
\providecommand \urlprefix  [0]{URL }%
\providecommand \Eprint [0]{\href }%
\providecommand \doibase [0]{http://dx.doi.org/}%
\providecommand \selectlanguage [0]{\@gobble}%
\providecommand \bibinfo  [0]{\@secondoftwo}%
\providecommand \bibfield  [0]{\@secondoftwo}%
\providecommand \translation [1]{[#1]}%
\providecommand \BibitemOpen [0]{}%
\providecommand \bibitemStop [0]{}%
\providecommand \bibitemNoStop [0]{.\EOS\space}%
\providecommand \EOS [0]{\spacefactor3000\relax}%
\providecommand \BibitemShut  [1]{\csname bibitem#1\endcsname}%
\let\auto@bib@innerbib\@empty
\bibitem [{\citenamefont {Ji}(2013)}]{Ji:2013dva}%
  \BibitemOpen
  \bibfield  {author} {\bibinfo {author} {\bibfnamefont {X.}~\bibnamefont
  {Ji}},\ }\href {\doibase 10.1103/PhysRevLett.110.262002} {\bibfield
  {journal} {\bibinfo  {journal} {Phys. Rev. Lett.}\ }\textbf {\bibinfo
  {volume} {110}},\ \bibinfo {pages} {262002} (\bibinfo {year} {2013})},\
  \Eprint {http://arxiv.org/abs/1305.1539} {arXiv:1305.1539 [hep-ph]}
  \BibitemShut {NoStop}%
\bibitem [{\citenamefont {Thomas}\ \emph {et~al.}(1981)\citenamefont {Thomas},
  \citenamefont {Theberge},\ and\ \citenamefont {Miller}}]{Thomas:1981vc}%
  \BibitemOpen
  \bibfield  {author} {\bibinfo {author} {\bibfnamefont {A.~W.}\ \bibnamefont
  {Thomas}}, \bibinfo {author} {\bibfnamefont {S.}~\bibnamefont {Theberge}}, \
  and\ \bibinfo {author} {\bibfnamefont {G.~A.}\ \bibnamefont {Miller}},\
  }\href {\doibase 10.1103/PhysRevD.24.216} {\bibfield  {journal} {\bibinfo
  {journal} {Phys. Rev.}\ }\textbf {\bibinfo {volume} {D24}},\ \bibinfo {pages}
  {216} (\bibinfo {year} {1981})}\BibitemShut {NoStop}%
\bibitem [{\citenamefont {Peng}\ \emph {et~al.}(1998)\citenamefont {Peng} \emph
  {et~al.}}]{Peng:1998pa}%
  \BibitemOpen
  \bibfield  {author} {\bibinfo {author} {\bibfnamefont {J.~C.}\ \bibnamefont
  {Peng}} \emph {et~al.} (\bibinfo {collaboration} {NuSea}),\ }\href {\doibase
  10.1103/PhysRevD.58.092004} {\bibfield  {journal} {\bibinfo  {journal} {Phys.
  Rev.}\ }\textbf {\bibinfo {volume} {D58}},\ \bibinfo {pages} {092004}
  (\bibinfo {year} {1998})},\ \Eprint {http://arxiv.org/abs/hep-ph/9804288}
  {arXiv:hep-ph/9804288 [hep-ph]} \BibitemShut {NoStop}%
\bibitem [{\citenamefont {Annand}\ \emph {et~al.}()\citenamefont {Annand},
  \citenamefont {Dutta}, \citenamefont {Keppel}, \citenamefont {King},\ and\
  \citenamefont {Wojtsekhowski~(spokespersons)}}]{Annand:2015}%
  \BibitemOpen
  \bibfield  {author} {\bibinfo {author} {\bibfnamefont {J.}~\bibnamefont
  {Annand}}, \bibinfo {author} {\bibfnamefont {D.}~\bibnamefont {Dutta}},
  \bibinfo {author} {\bibfnamefont {C.}~\bibnamefont {Keppel}}, \bibinfo
  {author} {\bibfnamefont {P.}~\bibnamefont {King}}, \ and\ \bibinfo {author}
  {\bibfnamefont {B.}~\bibnamefont {Wojtsekhowski~(spokespersons)}},\
  }\href@noop {} {\bibinfo  {journal} {Jefferson Lab Experiment PR12-15-006}\
  }\BibitemShut {NoStop}%
\bibitem [{\citenamefont {Szczepaniak}\ \emph {et~al.}(1994)\citenamefont
  {Szczepaniak}, \citenamefont {Ji},\ and\ \citenamefont
  {Cotanch}}]{Szczepaniak:1993uq}%
  \BibitemOpen
\bibfield  {journal} {  }\bibfield  {author} {\bibinfo {author} {\bibfnamefont
  {A.}~\bibnamefont {Szczepaniak}}, \bibinfo {author} {\bibfnamefont {C.-R.}\
  \bibnamefont {Ji}}, \ and\ \bibinfo {author} {\bibfnamefont {S.~R.}\
  \bibnamefont {Cotanch}},\ }\href {\doibase 10.1103/PhysRevD.49.3466}
  {\bibfield  {journal} {\bibinfo  {journal} {Phys. Rev.}\ }\textbf {\bibinfo
  {volume} {D49}},\ \bibinfo {pages} {3466} (\bibinfo {year} {1994})},\ \Eprint
  {http://arxiv.org/abs/hep-ph/9309284} {arXiv:hep-ph/9309284 [hep-ph]}
  \BibitemShut {NoStop}%
\bibitem [{\citenamefont {Frederico}\ and\ \citenamefont
  {Miller}(1994)}]{Frederico:1994dx}%
  \BibitemOpen
  \bibfield  {author} {\bibinfo {author} {\bibfnamefont {T.}~\bibnamefont
  {Frederico}}\ and\ \bibinfo {author} {\bibfnamefont {G.~A.}\ \bibnamefont
  {Miller}},\ }\href {\doibase 10.1103/PhysRevD.50.210} {\bibfield  {journal}
  {\bibinfo  {journal} {Phys. Rev.}\ }\textbf {\bibinfo {volume} {D50}},\
  \bibinfo {pages} {210} (\bibinfo {year} {1994})}\BibitemShut {NoStop}%
\bibitem [{\citenamefont {Bentz}\ \emph {et~al.}(1999)\citenamefont {Bentz},
  \citenamefont {Hama}, \citenamefont {Matsuki},\ and\ \citenamefont
  {Yazaki}}]{Bentz:1999gx}%
  \BibitemOpen
  \bibfield  {author} {\bibinfo {author} {\bibfnamefont {W.}~\bibnamefont
  {Bentz}}, \bibinfo {author} {\bibfnamefont {T.}~\bibnamefont {Hama}},
  \bibinfo {author} {\bibfnamefont {T.}~\bibnamefont {Matsuki}}, \ and\
  \bibinfo {author} {\bibfnamefont {K.}~\bibnamefont {Yazaki}},\ }\href
  {\doibase 10.1016/S0375-9474(99)00130-X} {\bibfield  {journal} {\bibinfo
  {journal} {Nucl. Phys.}\ }\textbf {\bibinfo {volume} {A651}},\ \bibinfo
  {pages} {143} (\bibinfo {year} {1999})},\ \Eprint
  {http://arxiv.org/abs/hep-ph/9901377} {arXiv:hep-ph/9901377 [hep-ph]}
  \BibitemShut {NoStop}%
\bibitem [{\citenamefont {Hecht}\ \emph {et~al.}(2001)\citenamefont {Hecht},
  \citenamefont {Roberts},\ and\ \citenamefont {Schmidt}}]{Hecht:2000xa}%
  \BibitemOpen
  \bibfield  {author} {\bibinfo {author} {\bibfnamefont {M.~B.}\ \bibnamefont
  {Hecht}}, \bibinfo {author} {\bibfnamefont {C.~D.}\ \bibnamefont {Roberts}},
  \ and\ \bibinfo {author} {\bibfnamefont {S.~M.}\ \bibnamefont {Schmidt}},\
  }\href {\doibase 10.1103/PhysRevC.63.025213} {\bibfield  {journal} {\bibinfo
  {journal} {Phys. Rev.}\ }\textbf {\bibinfo {volume} {C63}},\ \bibinfo {pages}
  {025213} (\bibinfo {year} {2001})},\ \Eprint
  {http://arxiv.org/abs/nucl-th/0008049} {arXiv:nucl-th/0008049 [nucl-th]}
  \BibitemShut {NoStop}%
\bibitem [{\citenamefont {Ji}\ \emph {et~al.}(2005)\citenamefont {Ji},
  \citenamefont {Ma},\ and\ \citenamefont {Yuan}}]{Ji:2004hz}%
  \BibitemOpen
  \bibfield  {author} {\bibinfo {author} {\bibfnamefont {X.-d.}\ \bibnamefont
  {Ji}}, \bibinfo {author} {\bibfnamefont {J.-P.}\ \bibnamefont {Ma}}, \ and\
  \bibinfo {author} {\bibfnamefont {F.}~\bibnamefont {Yuan}},\ }\href {\doibase
  10.1016/j.physletb.2005.02.019} {\bibfield  {journal} {\bibinfo  {journal}
  {Phys. Lett.}\ }\textbf {\bibinfo {volume} {B610}},\ \bibinfo {pages} {247}
  (\bibinfo {year} {2005})},\ \Eprint {http://arxiv.org/abs/hep-ph/0411382}
  {arXiv:hep-ph/0411382 [hep-ph]} \BibitemShut {NoStop}%
\bibitem [{\citenamefont {Chen}\ \emph {et~al.}(2017)\citenamefont {Chen},
  \citenamefont {Ji},\ and\ \citenamefont {Zhang}}]{Chen:2016fxx}%
  \BibitemOpen
  \bibfield  {author} {\bibinfo {author} {\bibfnamefont {J.-W.}\ \bibnamefont
  {Chen}}, \bibinfo {author} {\bibfnamefont {X.}~\bibnamefont {Ji}}, \ and\
  \bibinfo {author} {\bibfnamefont {J.-H.}\ \bibnamefont {Zhang}},\ }\href
  {\doibase 10.1016/j.nuclphysb.2016.12.004} {\bibfield  {journal} {\bibinfo
  {journal} {Nucl. Phys.}\ }\textbf {\bibinfo {volume} {B915}},\ \bibinfo
  {pages} {1} (\bibinfo {year} {2017})},\ \Eprint
  {http://arxiv.org/abs/1609.08102} {arXiv:1609.08102 [hep-ph]} \BibitemShut
  {NoStop}%
\bibitem [{\citenamefont
  {Radyushkin}(2017{\natexlab{a}})}]{Radyushkin:2016hsy}%
  \BibitemOpen
  \bibfield  {author} {\bibinfo {author} {\bibfnamefont {A.}~\bibnamefont
  {Radyushkin}},\ }\href {\doibase 10.1016/j.physletb.2017.02.019} {\bibfield
  {journal} {\bibinfo  {journal} {Phys. Lett.}\ }\textbf {\bibinfo {volume}
  {B767}},\ \bibinfo {pages} {314} (\bibinfo {year} {2017}{\natexlab{a}})},\
  \Eprint {http://arxiv.org/abs/1612.05170} {arXiv:1612.05170 [hep-ph]}
  \BibitemShut {NoStop}%
\bibitem [{\citenamefont
  {Radyushkin}(2017{\natexlab{b}})}]{Radyushkin:2017ffo}%
  \BibitemOpen
  \bibfield  {author} {\bibinfo {author} {\bibfnamefont {A.}~\bibnamefont
  {Radyushkin}},\ }\href {\doibase 10.1016/j.physletb.2017.05.024} {\bibfield
  {journal} {\bibinfo  {journal} {Phys. Lett.}\ }\textbf {\bibinfo {volume}
  {B770}},\ \bibinfo {pages} {514} (\bibinfo {year} {2017}{\natexlab{b}})},\
  \Eprint {http://arxiv.org/abs/1702.01726} {arXiv:1702.01726 [hep-ph]}
  \BibitemShut {NoStop}%
\bibitem [{\citenamefont {Carlson}\ and\ \citenamefont
  {Freid}(2017)}]{Carlson:2017gpk}%
  \BibitemOpen
  \bibfield  {author} {\bibinfo {author} {\bibfnamefont {C.~E.}\ \bibnamefont
  {Carlson}}\ and\ \bibinfo {author} {\bibfnamefont {M.}~\bibnamefont
  {Freid}},\ }\href {\doibase 10.1103/PhysRevD.95.094504} {\bibfield  {journal}
  {\bibinfo  {journal} {Phys. Rev.}\ }\textbf {\bibinfo {volume} {D95}},\
  \bibinfo {pages} {094504} (\bibinfo {year} {2017})},\ \Eprint
  {http://arxiv.org/abs/1702.05775} {arXiv:1702.05775 [hep-ph]} \BibitemShut
  {NoStop}%
\bibitem [{\citenamefont {Lin}\ \emph {et~al.}(2015)\citenamefont {Lin},
  \citenamefont {Chen}, \citenamefont {Cohen},\ and\ \citenamefont
  {Ji}}]{Lin:2014zya}%
  \BibitemOpen
  \bibfield  {author} {\bibinfo {author} {\bibfnamefont {H.-W.}\ \bibnamefont
  {Lin}}, \bibinfo {author} {\bibfnamefont {J.-W.}\ \bibnamefont {Chen}},
  \bibinfo {author} {\bibfnamefont {S.~D.}\ \bibnamefont {Cohen}}, \ and\
  \bibinfo {author} {\bibfnamefont {X.}~\bibnamefont {Ji}},\ }\href {\doibase
  10.1103/PhysRevD.91.054510} {\bibfield  {journal} {\bibinfo  {journal} {Phys.
  Rev.}\ }\textbf {\bibinfo {volume} {D91}},\ \bibinfo {pages} {054510}
  (\bibinfo {year} {2015})},\ \Eprint {http://arxiv.org/abs/1402.1462}
  {arXiv:1402.1462 [hep-ph]} \BibitemShut {NoStop}%
\bibitem [{\citenamefont {Chen}\ \emph {et~al.}(2016)\citenamefont {Chen},
  \citenamefont {Cohen}, \citenamefont {Ji}, \citenamefont {Lin},\ and\
  \citenamefont {Zhang}}]{Chen:2016utp}%
  \BibitemOpen
  \bibfield  {author} {\bibinfo {author} {\bibfnamefont {J.-W.}\ \bibnamefont
  {Chen}}, \bibinfo {author} {\bibfnamefont {S.~D.}\ \bibnamefont {Cohen}},
  \bibinfo {author} {\bibfnamefont {X.}~\bibnamefont {Ji}}, \bibinfo {author}
  {\bibfnamefont {H.-W.}\ \bibnamefont {Lin}}, \ and\ \bibinfo {author}
  {\bibfnamefont {J.-H.}\ \bibnamefont {Zhang}},\ }\href {\doibase
  10.1016/j.nuclphysb.2016.07.033} {\bibfield  {journal} {\bibinfo  {journal}
  {Nucl. Phys.}\ }\textbf {\bibinfo {volume} {B911}},\ \bibinfo {pages} {246}
  (\bibinfo {year} {2016})},\ \Eprint {http://arxiv.org/abs/1603.06664}
  {arXiv:1603.06664 [hep-ph]} \BibitemShut {NoStop}%
\bibitem [{\citenamefont {Gamberg}\ \emph {et~al.}(2015)\citenamefont
  {Gamberg}, \citenamefont {Kang}, \citenamefont {Vitev},\ and\ \citenamefont
  {Xing}}]{Gamberg:2014zwa}%
  \BibitemOpen
  \bibfield  {author} {\bibinfo {author} {\bibfnamefont {L.}~\bibnamefont
  {Gamberg}}, \bibinfo {author} {\bibfnamefont {Z.-B.}\ \bibnamefont {Kang}},
  \bibinfo {author} {\bibfnamefont {I.}~\bibnamefont {Vitev}}, \ and\ \bibinfo
  {author} {\bibfnamefont {H.}~\bibnamefont {Xing}},\ }\href {\doibase
  10.1016/j.physletb.2015.02.021} {\bibfield  {journal} {\bibinfo  {journal}
  {Phys. Lett.}\ }\textbf {\bibinfo {volume} {B743}},\ \bibinfo {pages} {112}
  (\bibinfo {year} {2015})},\ \Eprint {http://arxiv.org/abs/1412.3401}
  {arXiv:1412.3401 [hep-ph]} \BibitemShut {NoStop}%
\bibitem [{\citenamefont
  {Radyushkin}(2017{\natexlab{c}})}]{Radyushkin:2017gjd}%
  \BibitemOpen
  \bibfield  {author} {\bibinfo {author} {\bibfnamefont {A.~V.}\ \bibnamefont
  {Radyushkin}},\ }\href {\doibase 10.1103/PhysRevD.95.056020} {\bibfield
  {journal} {\bibinfo  {journal} {Phys. Rev.}\ }\textbf {\bibinfo {volume}
  {D95}},\ \bibinfo {pages} {056020} (\bibinfo {year} {2017}{\natexlab{c}})},\
  \Eprint {http://arxiv.org/abs/1701.02688} {arXiv:1701.02688 [hep-ph]}
  \BibitemShut {NoStop}%
\bibitem [{\citenamefont {Broniowski}\ and\ \citenamefont
  {Ruiz~Arriola}(2017)}]{Broniowski:2017wbr}%
  \BibitemOpen
  \bibfield  {author} {\bibinfo {author} {\bibfnamefont {W.}~\bibnamefont
  {Broniowski}}\ and\ \bibinfo {author} {\bibfnamefont {E.}~\bibnamefont
  {Ruiz~Arriola}},\ }\href@noop {} {\  (\bibinfo {year} {2017})},\ \Eprint
  {http://arxiv.org/abs/1707.09588} {arXiv:1707.09588 [hep-ph]} \BibitemShut
  {NoStop}%
\bibitem [{\citenamefont {Nam}(2017)}]{Nam:2017gzm}%
  \BibitemOpen
  \bibfield  {author} {\bibinfo {author} {\bibfnamefont {S.-i.}\ \bibnamefont
  {Nam}},\ }\href@noop {} {\  (\bibinfo {year} {2017})},\ \Eprint
  {http://arxiv.org/abs/1704.03824} {arXiv:1704.03824 [hep-ph]} \BibitemShut
  {NoStop}%
\bibitem [{\citenamefont {Zhang}\ \emph {et~al.}(2017)\citenamefont {Zhang},
  \citenamefont {Chen}, \citenamefont {Ji}, \citenamefont {Jin},\ and\
  \citenamefont {Lin}}]{Zhang:2017bzy}%
  \BibitemOpen
  \bibfield  {author} {\bibinfo {author} {\bibfnamefont {J.-H.}\ \bibnamefont
  {Zhang}}, \bibinfo {author} {\bibfnamefont {J.-W.}\ \bibnamefont {Chen}},
  \bibinfo {author} {\bibfnamefont {X.}~\bibnamefont {Ji}}, \bibinfo {author}
  {\bibfnamefont {L.}~\bibnamefont {Jin}}, \ and\ \bibinfo {author}
  {\bibfnamefont {H.-W.}\ \bibnamefont {Lin}},\ }\href {\doibase
  10.1103/PhysRevD.95.094514} {\bibfield  {journal} {\bibinfo  {journal} {Phys.
  Rev.}\ }\textbf {\bibinfo {volume} {D95}},\ \bibinfo {pages} {094514}
  (\bibinfo {year} {2017})},\ \Eprint {http://arxiv.org/abs/1702.00008}
  {arXiv:1702.00008 [hep-lat]} \BibitemShut {NoStop}%
\bibitem [{\citenamefont {Kang}\ and\ \citenamefont {Qiu}(2009)}]{Kang:2008ey}%
  \BibitemOpen
  \bibfield  {author} {\bibinfo {author} {\bibfnamefont {Z.-B.}\ \bibnamefont
  {Kang}}\ and\ \bibinfo {author} {\bibfnamefont {J.-W.}\ \bibnamefont {Qiu}},\
  }\href {\doibase 10.1103/PhysRevD.79.016003} {\bibfield  {journal} {\bibinfo
  {journal} {Phys. Rev.}\ }\textbf {\bibinfo {volume} {D79}},\ \bibinfo {pages}
  {016003} (\bibinfo {year} {2009})},\ \Eprint {http://arxiv.org/abs/0811.3101}
  {arXiv:0811.3101 [hep-ph]} \BibitemShut {NoStop}%
\bibitem [{\citenamefont {Collins}(2013)}]{Collins:2011zzd}%
  \BibitemOpen
  \bibfield  {author} {\bibinfo {author} {\bibfnamefont {J.}~\bibnamefont
  {Collins}},\ }\href {http://www.cambridge.org/de/knowledge/isbn/item5756723}
  {\emph {\bibinfo {title} {{Foundations of perturbative QCD}}}}\ (\bibinfo
  {publisher} {Cambridge University Press},\ \bibinfo {year}
  {2013})\BibitemShut {NoStop}%
\bibitem [{\citenamefont {Best}\ \emph {et~al.}(1997)\citenamefont {Best},
  \citenamefont {Gockeler}, \citenamefont {Horsley}, \citenamefont
  {Ilgenfritz}, \citenamefont {Perlt}, \citenamefont {Rakow}, \citenamefont
  {Schafer}, \citenamefont {Schierholz}, \citenamefont {Schiller},\ and\
  \citenamefont {Schramm}}]{Best:1997qp}%
  \BibitemOpen
  \bibfield  {author} {\bibinfo {author} {\bibfnamefont {C.}~\bibnamefont
  {Best}}, \bibinfo {author} {\bibfnamefont {M.}~\bibnamefont {Gockeler}},
  \bibinfo {author} {\bibfnamefont {R.}~\bibnamefont {Horsley}}, \bibinfo
  {author} {\bibfnamefont {E.-M.}\ \bibnamefont {Ilgenfritz}}, \bibinfo
  {author} {\bibfnamefont {H.}~\bibnamefont {Perlt}}, \bibinfo {author}
  {\bibfnamefont {P.~E.~L.}\ \bibnamefont {Rakow}}, \bibinfo {author}
  {\bibfnamefont {A.}~\bibnamefont {Schafer}}, \bibinfo {author} {\bibfnamefont
  {G.}~\bibnamefont {Schierholz}}, \bibinfo {author} {\bibfnamefont
  {A.}~\bibnamefont {Schiller}}, \ and\ \bibinfo {author} {\bibfnamefont
  {S.}~\bibnamefont {Schramm}},\ }\href {\doibase 10.1103/PhysRevD.56.2743}
  {\bibfield  {journal} {\bibinfo  {journal} {Phys. Rev.}\ }\textbf {\bibinfo
  {volume} {D56}},\ \bibinfo {pages} {2743} (\bibinfo {year} {1997})},\ \Eprint
  {http://arxiv.org/abs/hep-lat/9703014} {arXiv:hep-lat/9703014 [hep-lat]}
  \BibitemShut {NoStop}%
\bibitem [{\citenamefont {Conway}\ \emph {et~al.}(1989)\citenamefont {Conway}
  \emph {et~al.}}]{Conway:1989fs}%
  \BibitemOpen
  \bibfield  {author} {\bibinfo {author} {\bibfnamefont {J.~S.}\ \bibnamefont
  {Conway}} \emph {et~al.},\ }\href {\doibase 10.1103/PhysRevD.39.92}
  {\bibfield  {journal} {\bibinfo  {journal} {Phys. Rev.}\ }\textbf {\bibinfo
  {volume} {D39}},\ \bibinfo {pages} {92} (\bibinfo {year} {1989})}\BibitemShut
  {NoStop}%
\bibitem [{\citenamefont {Gluck}\ \emph {et~al.}(1992)\citenamefont {Gluck},
  \citenamefont {Reya},\ and\ \citenamefont {Vogt}}]{Gluck:1991ey}%
  \BibitemOpen
  \bibfield  {author} {\bibinfo {author} {\bibfnamefont {M.}~\bibnamefont
  {Gluck}}, \bibinfo {author} {\bibfnamefont {E.}~\bibnamefont {Reya}}, \ and\
  \bibinfo {author} {\bibfnamefont {A.}~\bibnamefont {Vogt}},\ }\href {\doibase
  10.1007/BF01559743} {\bibfield  {journal} {\bibinfo  {journal} {Z. Phys.}\
  }\textbf {\bibinfo {volume} {C53}},\ \bibinfo {pages} {651} (\bibinfo {year}
  {1992})}\BibitemShut {NoStop}%
\bibitem [{\citenamefont {Pumplin}(2006)}]{Pumplin:2005yf}%
  \BibitemOpen
  \bibfield  {author} {\bibinfo {author} {\bibfnamefont {J.}~\bibnamefont
  {Pumplin}},\ }\href {\doibase 10.1103/PhysRevD.73.114015} {\bibfield
  {journal} {\bibinfo  {journal} {Phys. Rev.}\ }\textbf {\bibinfo {volume}
  {D73}},\ \bibinfo {pages} {114015} (\bibinfo {year} {2006})},\ \Eprint
  {http://arxiv.org/abs/hep-ph/0508184} {arXiv:hep-ph/0508184 [hep-ph]}
  \BibitemShut {NoStop}%
\bibitem [{\citenamefont {Hobbs}\ \emph {et~al.}(2014)\citenamefont {Hobbs},
  \citenamefont {Londergan},\ and\ \citenamefont
  {Melnitchouk}}]{Hobbs:2013bia}%
  \BibitemOpen
  \bibfield  {author} {\bibinfo {author} {\bibfnamefont {T.~J.}\ \bibnamefont
  {Hobbs}}, \bibinfo {author} {\bibfnamefont {J.~T.}\ \bibnamefont
  {Londergan}}, \ and\ \bibinfo {author} {\bibfnamefont {W.}~\bibnamefont
  {Melnitchouk}},\ }\href {\doibase 10.1103/PhysRevD.89.074008} {\bibfield
  {journal} {\bibinfo  {journal} {Phys. Rev.}\ }\textbf {\bibinfo {volume}
  {D89}},\ \bibinfo {pages} {074008} (\bibinfo {year} {2014})},\ \Eprint
  {http://arxiv.org/abs/1311.1578} {arXiv:1311.1578 [hep-ph]} \BibitemShut
  {NoStop}%
\end{thebibliography}
%
%
%
\end{document}